\documentclass[sigconf]{acmart}
\usepackage{enumitem}

\AtBeginDocument{%
  }

\usepackage{subcaption}
\setcopyright{acmlicensed}
\copyrightyear{2026}
\acmYear{2026}
\acmDOI{XXXXXXX.XXXXXXX}
\acmConference[KDD '26]{The 32nd ACM SIGKDD Conference on Knowledge Discovery and Data Mining, AI for Sciences Track}{August 9--13, 2026}{Jeju, Korea}
\acmISBN{978-1-4503-XXXX-X/2026/08}

\usepackage[table]{xcolor}
\definecolor{highlightblue}{RGB}{235, 240, 255}
\definecolor{highlightred}{RGB}{255, 240, 240}
\definecolor{darkgreen}{RGB}{0, 100, 0}
\definecolor{darkred}{RGB}{139, 0, 0}

\definecolor{stanfordred}{HTML}{8C1515}
\definecolor{superlightgray}{RGB}{225, 225, 225}
\definecolor{lightblue}{RGB}{230, 240, 250}

\begin{document}

\title{Neural Gaussian Radio Fields for Channel Estimation}

\author{Muhammad Umer}
\authornote{These authors contributed equally to this research.}
\orcid{0009-0001-8751-6100}
\email{mumer@stanford.edu}
\affiliation{%
  \institution{Stanford University}
  \city{Stanford}
  \state{California}
  \country{USA}
}

\author{Muhammad Ahmed Mohsin}
\authornotemark[1]
\orcid{0009-0005-2766-0345}
\email{muahmed@stanford.edu}
\affiliation{%
  \institution{Stanford University}
  \city{Stanford}
  \state{California}
  \country{USA}
}

\author{Ahsan Bilal}
\authornotemark[1]
\orcid{0009-0002-7044-9316}
\email{ahsan.bilal-1@ou.edu}
\affiliation{%
  \institution{University of Oklahoma}
  \city{Norman}
  \state{Oklahoma}
  \country{USA}
}

\author{John M. Cioffi}
\orcid{0000-0003-1353-8101}
\email{jcioffi@stanford.edu}
\affiliation{%
  \institution{Stanford University}
  \city{Stanford}
  \state{California}
  \country{USA}
}

\renewcommand{\shortauthors}{Umer et al.}

\begin{abstract}
  Accurate channel state information (CSI) is a critical bottleneck in modern wireless networks, with pilot overhead consuming 11\% to 21\% of transmission bandwidth and feedback delays causing severe throughput degradation under mobility. Addressing this requires rethinking how neural fields represent coherent wave phenomena. This work introduces \textit{neural Gaussian radio fields (\textcolor{stanfordred}{nGRF})}, a physics-informed framework that fundamentally reframes neural field design by replacing view-dependent rasterization with direct complex-valued aggregation in 3D space. This approach natively models wave superposition rather than visual occlusion. The architectural shift transforms the learning objective from function-fitting to source-recovery, a well-posed inverse problem grounded in electromagnetic theory. While demonstrated for wireless channel estimation, the core principle of explicit primitive-based fields with physics-constrained aggregation extends naturally to any coherent wave-based domain, including acoustic propagation, seismic imaging, and ultrasound reconstruction. Evaluations show that the inductive bias of \textcolor{stanfordred}{nGRF} achieves 10.9 dB higher prediction SNR than state-of-the-art methods with 220$\times$ faster inference (1.1 ms vs. 242 ms), 18$\times$ lower measurement density, and 180$\times$ faster training. For large-scale outdoor environments where implicit methods fail, \textcolor{stanfordred}{nGRF} achieves 28.32 dB SNR, demonstrating that structured representations supplemented by domain physics can fundamentally outperform generic deep learning architectures.
\end{abstract}


\begin{teaserfigure}
  \centering
  \includegraphics[width=\textwidth]{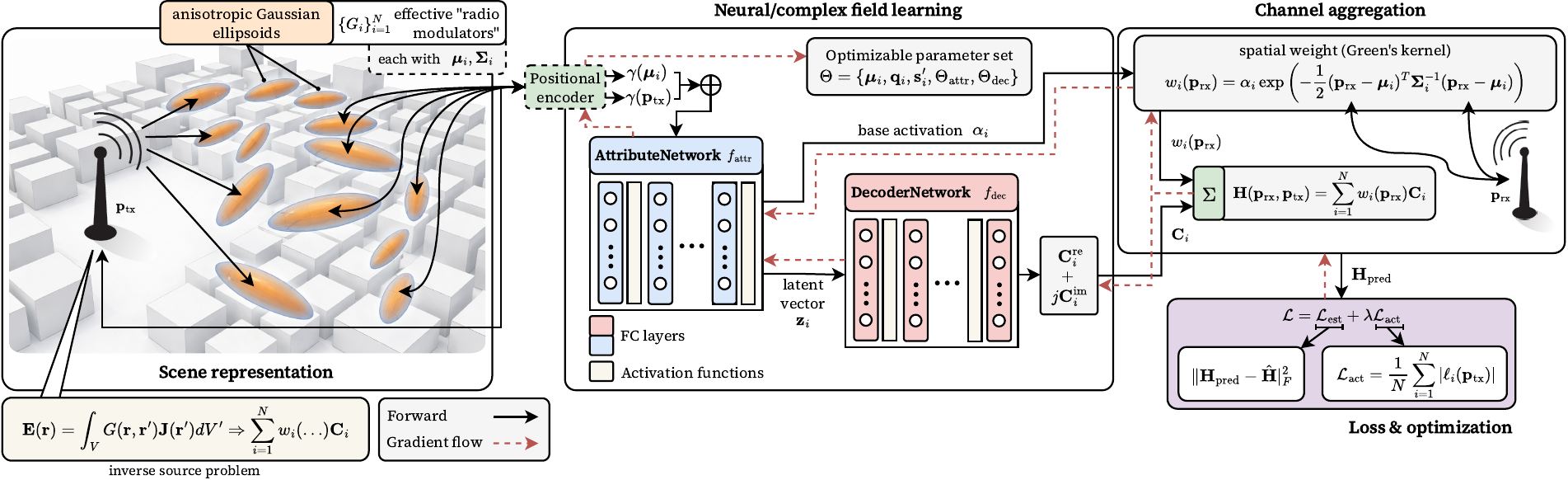}
  \caption{\textit{Overview of the \textcolor{stanfordred}{nGRF} framework.} Anisotropic 3D Gaussian primitives $\{G_i\}_{i=1}^N$ represent the radio environment as effective scattering sources (\textit{left}). The \texttt{AttributeNetwork} maps positionally encoded Gaussian centers and transmitter position to latent features $\mathbf{z}_i$ and activations $\alpha_i$; the \texttt{DecoderNetwork} produces complex-valued channel contributions $\mathbf{C}_i = \mathbf{C}_i^{\text{re}} + j\mathbf{C}_i^{\text{im}}$ (\textit{center}). Channel aggregation performs electromagnetic superposition; each contribution is weighted by a spatial kernel $w_i(\mathbf{p}_{\text{rx}})$ derived from the Mahalanobis distance to the receiver, and the full channel matrix $\mathbf{H}$ is obtained via coherent summation $\mathbf{H} = \sum_i w_i(\mathbf{p}_{\text{rx}})\,\mathbf{C}_i$ (\textit{right}). The entire pipeline can be thought of as a discretization of the electromagnetic integral equation $\mathbf{E}(\mathbf{r}) = \int_V G(\mathbf{r},\mathbf{r}')\,\mathbf{J}(\mathbf{r}')\,dV'$, which is trained end-to-end via gradient descent on a composite loss that combines Frobenius-norm error against reference channel estimates and activation sparsity (\textit{bottom right}).}
  \label{fig:teaser}
  \Description{A schematic pipeline of the neural Gaussian radio fields. From left to right: (1) A 3D environment is shown with a transmitter and multiple Gaussian primitives (ellipsoids) representing scattering sources. (2) A neural network block titled 'AttributeNetwork' takes the position of a Gaussian and the transmitter as input. (3) This feeds into a 'DecoderNetwork' which outputs complex matrices. (4) The final stage shows a 'channel aggregation' step where these matrices are summed together, weighted by spatial distance, to form the final channel matrix H.}
\end{teaserfigure}


\maketitle

\section{Introduction}
Despite decades of work, real-time channel state information (CSI) \emph{estimation and prediction} remains the principal unresolved bottleneck in both current and next-generation wireless networks. CSI, represented by the complex matrix $\mathbf{H}$, captures how signals propagate through direct paths, reflections, diffractions, and scattering between the transmitter and receiver. Accurate CSI enables transmitters to adapt their waveforms, power levels, and spatial precoding to channel conditions, directly determining achievable data rates and link reliability~\cite{bai2003error}. The sub-millisecond latencies and Gbps data rates targeted by 5G and 6G networks demand high-fidelity, low-overhead CSI estimation~\cite{giordani2020toward}.

Multiple-input multiple-output (MIMO) technology uses multiple antennas at the transmitter and receiver to send multiple parallel data streams over the same frequency band. This \emph{spatial multiplexing} boosts data rates and capacity without extra bandwidth. With \(N\) antennas at each end, the theoretical peak-throughput bound scales linearly by a factor of \(N\). Massive MIMO further scales antenna arrays by an order of magnitude, making $\mathbf{H}$ difficult to characterize~\cite{balevi2020massive}. Two fundamental challenges make CSI estimation and prediction difficult to resolve. First, pilot overhead is a significant bottleneck. Because every pilot resource element displaces data, pilots consume 11\%-21\% of 5G NR bandwidth~\cite{dahlman20145g, lin2022overview}, a proportion that grows further in high-mobility and cell-free massive MIMO settings~\cite{jardosh2005understanding}. Reducing pilots is essential for efficiency, yet doing so risks inaccurate CSI.

Second, even after channel information is acquired, it quickly becomes outdated in \emph{dynamic environments} due to channel aging~\cite{truong2013effects}. Under mobility, wireless channels decorrelate in milliseconds; at 3.5\,GHz with a user moving at ${\sim}$30\,km/h, the coherence time is only 2\,ms~\cite{wang2024robust}, so a 4\,ms feedback delay can cut the data rate by approximately 50\%. A mmWave channel can decorrelate within a single 1\,ms 5G subframe. Furthermore, dense networks face \emph{pilot contamination}, where inter-cell pilot reuse degrades CSI accuracy, a known bottleneck for MIMO systems~\cite{elijah2015comprehensive}.

Many AI-driven CSI estimators, however, disregard the physical 3D structure governing radio propagation. Data-driven methods, such as generative models or recurrent networks, treat the channel as an abstract data vector, lacking the inductive bias of the underlying physics and often incurring high latency from iterative sampling or large network backbones~\cite{arvinte2022mimo, aldossari2019machine}. Recent neural field approaches, while spatially aware, suffer from their own architectural limitations. Neural radiance field (NeRF)-based models rely on slow, implicit representations that require computationally expensive volumetric integration for every channel query~\cite{lu2024newrf, Zhao_2023}. Concurrently, methods adapting 3D Gaussian splatting (3DGS) for channel modeling are ill-suited for the task; they regress scalar power and apply 2D projections designed for visual rendering, failing to capture the complex-valued nature of electromagnetic fields~\cite{wen2025neural, niemeyer2025radsplat}. These limitations motivate a new modeling design that is both computationally efficient and physically grounded.

This work introduces \emph{neural Gaussian radio fields (\textcolor{stanfordred}{nGRF})}, a principled framework that fundamentally rethinks how neural fields represent coherent wave phenomena. Rather than adapting visual rendering techniques to radio propagation, \textcolor{stanfordred}{nGRF} is designed from first principles: it replaces view-dependent rasterization with physics-based complex-valued aggregation that directly implements electromagnetic superposition. Each 3D Gaussian primitive acts as a learned source in an inverse problem formulation, where channel estimation becomes source recovery rather than function fitting, transforming an ill-posed regression task into a well-conditioned electromagnetic inverse problem. This architectural choice, grounded in Green's function theory, provides three conceptual advantages: (1) \textit{physical consistency}, as the model natively respects wave superposition and phase coherence; (2) \textit{computational efficiency}, as explicit primitives eliminate costly volumetric integration; and (3) \textit{generalizability}, as the framework extends beyond wireless channels to any linear wave equation (acoustics, seismic, ultrasound) by substituting the appropriate Green's kernel while preserving the aggregation structure.

\textbf{Contributions.} \textbf{(i) Paradigm shift in neural field design.} \textcolor{stanfordred}{nGRF} demonstrates that \textit{physics-informed explicit representations fundamentally outperform generic implicit architectures} for coherent field phenomena. By directly modeling wave superposition through complex-valued 3D aggregation rather than alpha-compositing (designed for occlusion), \textcolor{stanfordred}{nGRF} reframes channel estimation from function fitting to source recovery as an inverse problem with provable stability. This design principle generalizes to any coherent wave domain governed by linear partial differential equations (PDEs). \textbf{(ii) State-of-the-art empirical performance.} Evaluations show 10.9 dB higher SNR than competing methods with 220$\times$ faster inference (1.1 ms), 18$\times$ lower measurement density, and 180$\times$ faster training, demonstrating that structured inductive biases can break the accuracy-efficiency trade-off that constrains generic networks. \textbf{(iii) Practical deployment viability.} For 5G NR systems, \textcolor{stanfordred}{nGRF} reduces pilot overhead from 11\%-21\% down to 0.2\% (96 bits of position data vs.\ thousands of resource elements), directly translating to increased spectral efficiency in operational networks. \textbf{(iv) Frequency-agnostic learning.} Training on a single subcarrier, \textcolor{stanfordred}{nGRF} generalizes across entire bands, providing evidence that it learns spatial propagation structure rather than frequency-specific mappings, enabling wideband prediction without per-subcarrier training. \textbf{(v) Scalability across regimes.} \textcolor{stanfordred}{nGRF} scales from indoor (10 m$^3$) to outdoor urban (1000 m$^3$) environments where implicit methods fail, achieving 28.32 dB SNR under 91\% NLOS conditions. This robust performance across 100$\times$ scene volume variation validates the generalization capacity of the physics-informed architecture.

\section{Related Work}
\label{sec:related_work}

Prior CSI estimation work spans three main approaches: data-driven models that disregard spatial physics, implicit neural fields that are computationally prohibitive, and explicit 3D Gaussian splatting (3DGS)-based methods that misapply visual rendering techniques to wave phenomena. The limitations of each motivate the need for a new approach.

\textbf{Data-driven approaches.}
Deep CSI estimators learn a direct pilot-to-CSI map with DNN/CNN/LSTM blocks~\cite{ye2018power,soltani2018dlce,le2021sensors_mlce,chen2021decnet}, which is fast but brittle across SNR/mobility/array shifts. Generative priors (VAE/GAN/diffusion) better capture channel statistics yet introduce heavy backbones and iterative sampling, and still treat CSI as an unstructured vector without spatial inductive bias~\cite{baur2022vae_mmse,baur2023vae_real,guo2024fedgan,arvinte2022score,zhou2025generative,jin2024gdm4mmimo}.

\textbf{Implicit neural fields.}
To inject spatial context into CSI prediction, recent work adapts NeRF-style implicit scene modeling to RF propagation, learning a continuous field from sparse measurements for channel prediction and coverage mapping (e.g., NeRF$^2$~\cite{Zhao_2023}, NeWRF~\cite{lu2024newrf}, SpecNeRF~\cite{basu2024specnerf}, and RIS-aware variants such as R-NeRF~\cite{yang2024rnerf}; see also follow-ups like NeRF-APT~\cite{shen2025nerfapt}). These methods inherit volumetric rendering~\cite{mildenhall2020nerf}: given a receiver ray $\mathbf{r}(t)$ with direction $\mathbf{d}$, the channel is obtained by integrating transmittance-weighted field contributions along the ray,
\begin{equation}
  \mathbf{H}(\mathbf{r},\mathbf{d})
  =\int_{t_{\mathrm{near}}}^{t_{\mathrm{far}}}
  T(t)\,\sigma\!\big(\mathbf{r}(t),\mathbf{d}\big)\,\mathbf{c}\!\big(\mathbf{r}(t),\mathbf{d}\big)\,dt,
\end{equation}
where $(\sigma,\mathbf{c}) = F_\Theta(\mathbf{r}(t),\mathbf{d})$, $\sigma(\cdot)$ models attenuation, $\mathbf{c}(\cdot)$ is the complex contribution, and $T(t)=\exp\!\big(-\int_{t_{\mathrm{near}}}^{t}\sigma(\mathbf{r}(s))\,ds\big)$ is transmittance. In practice, this integral is discretized into a sum over $N_s$ samples, requiring $N_s$ multilayer perceptron (MLP) evaluations per ray (and often many rays per query), so compute scales directly with sampling density and scene complexity. This implicit ray-marching is the primary bottleneck: even with general NeRF accelerations such as hash-grid encodings (Instant-NGP)~\cite{muller2022instantngp} or specialized ray-marching toolboxes (NerfAcc)~\cite{li2022nerfacc}, RF-NeRF inference typically remains far above millisecond PHY budgets, limiting real-time deployment despite good 3D generalization.

\textbf{3DGS-based methods.}
To overcome NeRF's ray-marching cost, recent RF works replace implicit fields with explicit Gaussian primitives for fast splatting-based rendering (e.g., RF-3DGS~\cite{gentile2024rf3dgs}, WRF-GS~\cite{wen2025neural}, RadSplatter~\cite{wang2025radsplatter}, and subsequent real-time RF Gaussian pipelines~\cite{zhang2025gsparc}). These methods inherit the graphics renderer of 3DGS~\cite{kerbl2023gaussian}, which projects 3D primitives onto the image plane and aggregates them via alpha-compositing:
\begin{equation}
  C = \sum_{i=1}^N c_i \alpha_i \prod_{j=1}^{i-1} (1 - \alpha_j),
\end{equation}
where Gaussians are depth-sorted and $\alpha_i$ models visibility and occlusion. This architectural bias is well-suited to \emph{radiance} but mismatched to \emph{electromagnetics}: RF propagation is governed by linear superposition of complex-valued fields rather than foreground occlusion, so alpha blending cannot represent interference, phase cancellation, or coherent MIMO structure. As a result, most 3DGS-RF systems target scalar radiomaps and path visualization~\cite{gentile2024rf3dgs,wang2025radsplatter}, but do not natively produce the full complex CSI required for coherent beamforming and MIMO detection. These limitations motivate an explicit yet physics-native model: \textcolor{stanfordred}{nGRF} retains explicit primitives for speed while replacing visibility-based compositing with complex field aggregation designed for wave superposition.

\section{Neural Gaussian Radio Fields}
\label{sec:ngrf}

\subsection{Problem description}
Channel estimation fundamentally requires solving Maxwell's equations for a given environment. For time-harmonic electromagnetic fields at frequency $\omega$, this reduces to solving the vector Helmholtz equation expressed as $\nabla \times \nabla \times \mathbf{E}(\mathbf{r}) - k^2 \mathbf{E}(\mathbf{r}) = -j\omega\mu_0 \mathbf{J}(\mathbf{r})$, where $\mathbf{E}(\mathbf{r})$ is the electric field at position $\mathbf{r} \in \mathbb{R}^3$, $k = \omega\sqrt{\mu_0\epsilon_0}$ is the wavenumber, and $\mathbf{J}(\mathbf{r})$ represents the current sources. Solving this partial differential equation (PDE) with the appropriate boundary conditions defined by the environment's geometry and materials would yield perfect CSI. However, this is computationally intractable for any non-trivial scene.

The solution to the Helmholtz equation can be expressed via a Green's function $G(\mathbf{r}, \mathbf{r}')$, which describes the field at $\mathbf{r}$ due to a point source at $\mathbf{r}'$. In free space, this is a spherical wave, $G_0(\mathbf{r}, \mathbf{r}') = e^{ik|\mathbf{r}-\mathbf{r}'|} / (4\pi|\mathbf{r}-\mathbf{r}'|)$, while in a complex environment, the total field is the sum of the incident field from the source and the scattered field from all interacting objects. The scattered field, in turn, can be described by an integral of the Green's function over the surfaces of all scatterers. Thus, the complex multipath channel results from the superposition of waves originating from a set of effective sources distributed throughout the environment.

This motivates representing the field using a basis of functions that can model these localized wave contributions. Under the paraxial approximation, where waves propagate primarily along a single direction, solutions to the Helmholtz equation take the form of Gaussian beams~\cite{liu2013gaussian}. Such a physical connection suggests that a superposition of anisotropic 3D Gaussian functions can serve as a flexible basis set for representing the electromagnetic field in its entirety.

Existing methods make different trade-offs. Ray tracing, for example, approximates the solution in the geometric optics limit (wavelength $\lambda \to 0$), treating waves as simple rays. This fails to capture wave phenomena like diffraction and sub-wavelength interference, which are needed for accurate channel modeling. Implicit neural fields based on NeRF~\cite{mildenhall2020nerf} attempt to learn a continuous volumetric representation of the field. However, these models lack physical priors for wave propagation. They are generic function approximators that must learn the field's structure from scratch, requiring dense measurements and computationally expensive volumetric integration to render a single channel estimate.

\textcolor{stanfordred}{nGRF} reframes electromagnetic field estimation as a structured inverse problem rather than generic function approximation. The key physical insight is that while fields are continuous, the scattering phenomena generating multipath (reflections, diffractions, material boundaries) are spatially localized. This motivates decomposing the total field $\mathbf{E}(\mathbf{r})$ into direct propagation and a discrete sum over effective scattering sources:
\begin{equation}
  \mathbf{E}(\mathbf{r}) \approx \mathbf{E}_{\text{LoS}}(\mathbf{r}) + \sum_{i=1}^N \mathcal{A}_i(\mathbf{p}_{\text{tx}}, \mathbf{p}_{\text{rx}}) G_i(\mathbf{r}; \boldsymbol{\mu}_i, \boldsymbol{\Sigma}_i),
\end{equation}
where each Gaussian primitive $G_i$ represents the spatial extent of the $i$-th scattering region with center $\boldsymbol{\mu}_i$ and anisotropic spread $\boldsymbol{\Sigma}_i$, while the complex amplitude $\mathcal{A}_i$ encodes transmitter-receiver-dependent excitation. This formulation is not an architectural convenience but a direct discretization of the electromagnetic integral equation $\mathbf{E}(\mathbf{r}) = \int_V G(\mathbf{r}, \mathbf{r}') \mathbf{J}(\mathbf{r}')\, dV'$, where $G_i$ approximates the Green's function propagator and $\mathcal{A}_i$ approximates the source distribution $\mathbf{J}$. Notably, this transforms channel estimation from ill-posed function fitting (learning arbitrary $\mathbf{H}: \mathbb{R}^6 \to \mathbb{C}^{N_t \times N_r}$) into conditioned source recovery, with finite parameters $\{\boldsymbol{\mu}_i, \boldsymbol{\Sigma}_i, \mathcal{A}_i\}_{i=1}^N$ of a physics-constrained basis. The learning objective shifts from minimizing prediction error to recovering the spatial structure of wave interactions, a problem with inherent regularization from the Gaussian basis and electromagnetic constraints, explaining \textcolor{stanfordred}{nGRF}'s superior data efficiency and generalization.

\subsection{nGRF design}

\textcolor{stanfordred}{nGRF} represents the radio environment as $N$ physics-constrained 3D Gaussian primitives $\{G_i\}_{i=1}^N$, each serving as a learned basis function in the source-recovery formulation. Unlike implicit neural fields that represent the environment as a black-box function $f_\theta: \mathbb{R}^6 \to \mathbb{C}^{N_t \times N_r}$, \textcolor{stanfordred}{nGRF}'s explicit primitives directly encode the spatial structure of scattering phenomena, providing interpretable representations of effective electromagnetic sources. Figure~\ref{fig:teaser} illustrates the overall architecture.


\textbf{Geometric parameterization.} Each Gaussian primitive $G_i$ is characterized by geometric properties that define its spatial influence: a mean position $\boldsymbol{\mu}_i \in \mathbb{R}^3$ specifying the scattering center location, and an anisotropic covariance $\boldsymbol{\Sigma}_i \in \mathbb{R}^{3 \times 3}$ encoding directional propagation characteristics. The unnormalized spatial density is
\begin{equation}
  G_i(\mathbf{x}) = \exp\left(-\frac{1}{2}(\mathbf{x}-\boldsymbol{\mu}_i)^T\boldsymbol{\Sigma}_i^{-1}(\mathbf{x}-\boldsymbol{\mu}_i)\right),
\end{equation}
where the Mahalanobis distance $(\mathbf{x}-\boldsymbol{\mu}_i)^T\boldsymbol{\Sigma}_i^{-1}(\mathbf{x}-\boldsymbol{\mu}_i)$ naturally implements distance-dependent attenuation with directional weighting. To maintain $\boldsymbol{\Sigma}_i \succ 0$ during gradient-based optimization, the covariance is factorized as $\boldsymbol{\Sigma}_i = \mathbf{R}_i \mathbf{S}_i \mathbf{S}_i^T \mathbf{R}_i^T$, where $\mathbf{R}_i \in \text{SO}(3)$ is a rotation (parameterized via unit quaternion $\mathbf{q}_i \in \mathbb{R}^4$) and $\mathbf{S}_i = \text{diag}(s_{i,1}, s_{i,2}, s_{i,3})$ contains scales enforced positive through $s_{i,j} = \exp(s'_{i,j})$. This parameterization ensures well-conditioned covariances while allowing the model to learn arbitrary ellipsoidal shapes and orientations.

\begin{figure}[t]
  \includegraphics[width=0.98\columnwidth]{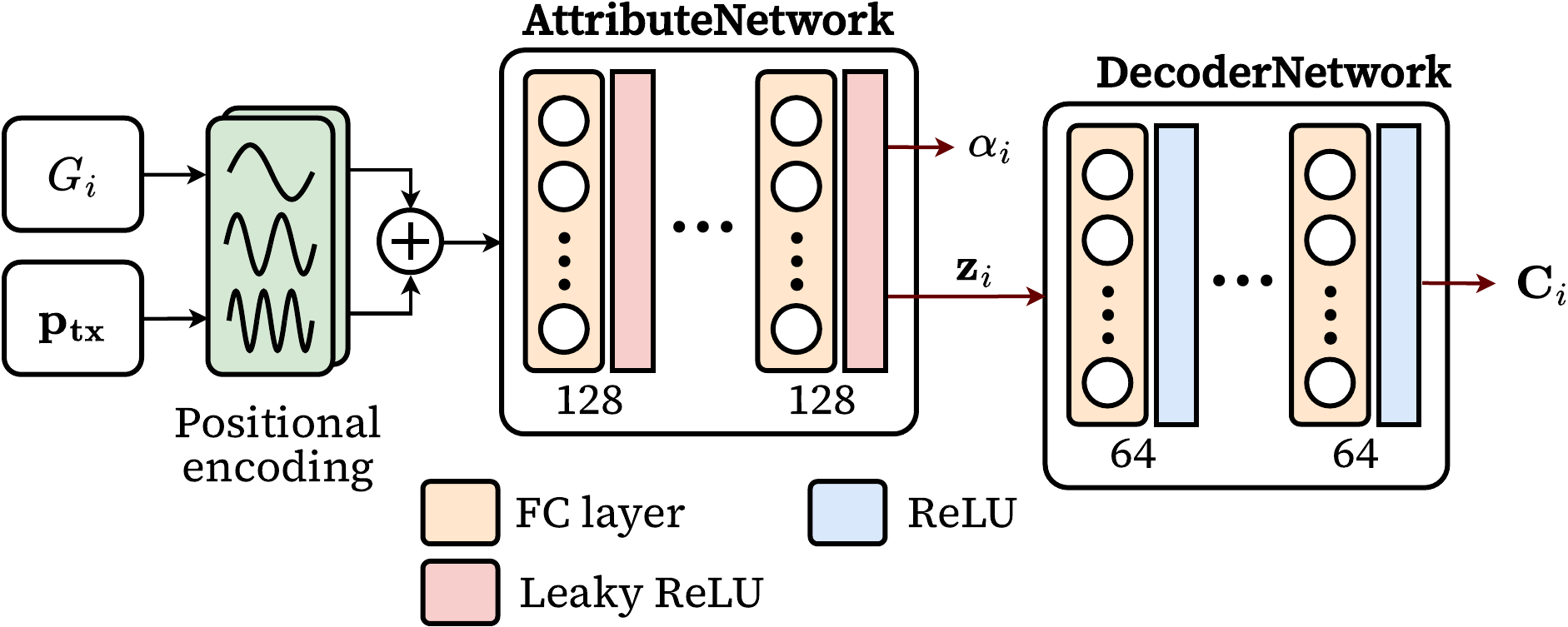}
  \caption{\textit{Neural architecture.} The \texttt{AttributeNetwork} processes Gaussian and transmitter positions through separate positional encoders to produce latent features and base activation logits. The \texttt{DecoderNetwork} then transforms the latent features into complex channel contributions.}
  \Description{A block diagram showing the data flow through two neural networks. The top section shows the 'AttributeNetwork', which takes positionally encoded vectors for Gaussian means and Transmitter positions as inputs, passing them through an MLP to output a latent feature vector 'z' and a scalar 'alpha'. The bottom section shows the 'DecoderNetwork', which takes the latent vector 'z', passes it through an MLP, and outputs two matrices representing the Real and Imaginary components of the channel contribution.}
  \label{fig:networks}
\end{figure}

\textbf{Electromagnetic attribute learning.} Each Gaussian's field contribution is not fixed but learned through two neural networks (with learnable parameters $\Theta_{\text{attr}}$ and $\Theta_{\text{dec}}$, respectively), as shown in Figure~\ref{fig:networks}, that map spatial context to complex-valued electromagnetic responses. This design separates \emph{where} scattering occurs (geometric parameters $\boldsymbol{\mu}_i, \boldsymbol{\Sigma}_i$) from \emph{how} it contributes to the channel, enabling the model to adapt to diverse propagation environments.

\texttt{AttributeNetwork} $f_{\text{attr}}: \mathbb{R}^{d_{\text{enc}}} \times \mathbb{R}^{d_{\text{enc}}} \to \mathbb{R}^d \times \mathbb{R}$ conditions each Gaussian's electromagnetic behavior on the transmitter location $\mathbf{p}_{\text{tx}}$, producing a latent feature vector $\mathbf{z}_i$ (encoding scattering characteristics such as polarization, phase response, and frequency dependence) and a scalar activation $\alpha_i$ (controlling contribution magnitude):
\begin{equation}
  (\mathbf{z}_i, \alpha_i) = f_{\text{attr}}(\gamma_L(\boldsymbol{\mu}_i), \gamma_L(\mathbf{p}_{\text{tx}}); \Theta_{\text{attr}}).
\end{equation}
Note that the inputs are transformed via multi-resolution positional encoding as in~\cite{mildenhall2020nerf}:
\begin{equation*}
  \gamma_L(\mathbf{x}) = \left[\mathbf{x}, \sin(2^0\pi\mathbf{x}), \cos(2^0\pi\mathbf{x}), \ldots, \sin(2^{L-1}\pi\mathbf{x}), \cos(2^{L-1}\pi\mathbf{x})\right],
\end{equation*}
which enables the network to capture rapid phase variations on the scale of the wavelength $\lambda$, essential for modeling interference patterns where sub-wavelength displacement causes destructive-to-constructive transitions. Without this encoding, standard MLPs struggle to represent the high-frequency spatial oscillations inherent to wave phenomena.

The \texttt{DecoderNetwork} $f_{\text{dec}}: \mathbb{R}^d \to \mathbb{C}^{N_t \times N_r}$ then maps the latent representation to the complex-valued MIMO channel contribution:
\begin{equation}
  \mathbf{C}_i = f_{\text{dec}}(\mathbf{z}_i; \Theta_{\text{dec}}) = \mathbf{C}_i^{\text{re}} + j \mathbf{C}_i^{\text{im}},
\end{equation}
where the real and imaginary components are output separately. This representation naturally encodes both magnitude and phase: $|\mathbf{C}_i|$ determines field strength while $\arg(\mathbf{C}_i)$ determines phase relationships necessary for coherent superposition. Together, the geometric parameters $(\boldsymbol{\mu}_i, \boldsymbol{\Sigma}_i)$ and learned electromagnetic attributes $(\alpha_i, \mathbf{z}_i \to \mathbf{C}_i)$ enable each Gaussian to function as an adaptive, transmitter-aware radio modulator, a localized source whose spatial extent, directional characteristics, and field contribution are jointly optimized to reconstruct the propagation environment.

\subsection{Channel rendering}
\label{sec:channel_rendering}
Channel rendering in \textcolor{stanfordred}{nGRF} implements wave superposition through direct aggregation, fundamentally departing from both volumetric integration and 2D alpha-compositing. This architectural choice is not merely a computational consideration; it follows directly from electromagnetic theory, wherein wave fields interact via linear superposition rather than geometric occlusion.

\textbf{Direct aggregation formulation.} For a transmitter-receiver pair $(\mathbf{p}_{\text{tx}}, \mathbf{p}_{\text{rx}})$, the MIMO channel matrix $\mathbf{H}(\mathbf{p}_{\text{rx}}, \mathbf{p}_{\text{tx}}) \in \mathbb{C}^{N_t \times N_r}$ is rendered as a weighted sum over all Gaussian primitives:
\begin{equation}
  \mathbf{H}(\mathbf{p}_{\text{rx}}, \mathbf{p}_{\text{tx}}) = \sum_{i=1}^N w_i(\mathbf{p}_{\text{rx}}) \cdot \mathbf{C}_i,
\end{equation}
where $\mathbf{C}_i \in \mathbb{C}^{N_t \times N_r}$ is the complex-valued field contribution from the $i$-th Gaussian, and $w_i \in \mathbb{R}_+$ is a spatial weight determining the influence of scattering region $i$ at receiver location $\mathbf{p}_{\text{rx}}$. This formulation directly discretizes the electromagnetic integral $\mathbf{E}(\mathbf{r}) = \int_V G(\mathbf{r}, \mathbf{r}') \mathbf{J}(\mathbf{r}')\, dV'$ introduced in Section~\ref{sec:ngrf}, where the integral over a continuous source distribution is approximated by a finite sum over localized sources, a standard technique in computational electromagnetics for solving scattering problems.

\textbf{Spatial weighting as propagation modeling.} The spatial weight implements distance-dependent and direction-dependent attenuation:
\begin{equation}
  w_i(\mathbf{p}_{\text{rx}}) = \alpha_i \cdot \exp\left(-\frac{1}{2}(\mathbf{p}_{\text{rx}} - \boldsymbol{\mu}_i)^T\boldsymbol{\Sigma}_i^{-1}(\mathbf{p}_{\text{rx}} - \boldsymbol{\mu}_i)\right),
\end{equation}
where $\alpha_i$ (base activation from the \texttt{AttributeNetwork}) controls the overall source strength, and the exponential term encodes spatial decay via the Mahalanobis distance $d_i^2 = (\mathbf{p}_{\text{rx}} - \boldsymbol{\mu}_i)^T\boldsymbol{\Sigma}_i^{-1}(\mathbf{p}_{\text{rx}} - \boldsymbol{\mu}_i)$. This formulation has three physical interpretations:

\textbf{(i) Distance-dependent decay.} Influence decreases with distance from the scattering center $\boldsymbol{\mu}_i$, analogous to free-space path loss but with learned, environment-specific characteristics rather than fixed $1/r^2$ scaling.

\textbf{(ii) Directional propagation.} The anisotropic covariance $\boldsymbol{\Sigma}_i$ enables direction-dependent weighting. Elongated Gaussians model phenomena like specular reflections from walls, where the field extends along the surface but decays rapidly perpendicular to it. This captures the angular selectivity of scattering without explicit ray tracing.

\textbf{(iii) Smooth spatial continuity.} The Gaussian form ensures $C^\infty$ differentiability with respect to $\mathbf{p}_{\text{rx}}$, which is required for gradient-based optimization and consistent with the physical expectation that small receiver displacements cause smooth (not discontinuous) channel variations outside nulls.

\textbf{Contrast with visual rendering.} This aggregation fundamentally differs from 3D Gaussian splatting methods adapted for radio~\cite{wen2025neural}, which employ alpha-compositing:
\begin{equation}
  C_{\text{2D}} = \sum_{i=1}^N c_i \alpha_i \prod_{j<i} (1 - \alpha_j). \quad \text{(occlusion-based, depth-sorted)}
\end{equation}
Alpha-compositing models geometric occlusion, where foreground objects block background ones, which is physically correct for light but fundamentally incorrect for electromagnetic waves that superimpose coherently regardless of relative depth. \textcolor{stanfordred}{nGRF}'s orderless summation $\mathbf{H} = \sum_i w_i \mathbf{C}_i$ implements wave superposition: all sources contribute to the total field, with constructive and destructive interference encoded in the complex phases of $\mathbf{C}_i$, not through multiplicative attenuation. This distinction is essential for capturing multipath phenomena where signals from behind an obstruction (via diffraction or scattering) still reach the receiver and interfere with line-of-sight components.

\subsection{Physical interpretation}
A fundamental distinction between \textcolor{stanfordred}{nGRF} and geometry-based rendering methods lies in how fields are composed. Visual rendering models occlusion, where foreground objects geometrically block background ones, whereas electromagnetic propagation obeys coherent superposition, where all fields contribute additively with phase-dependent interference. \textcolor{stanfordred}{nGRF}'s architecture explicitly separates these physical mechanisms: spatial weights $w_i(\mathbf{p}_{\text{rx}}) = \alpha_i \cdot \exp(-\frac{1}{2}(\mathbf{p}_{\text{rx}} - \boldsymbol{\mu}_i)^T \boldsymbol{\Sigma}_i^{-1}(\mathbf{p}_{\text{rx}} - \boldsymbol{\mu}_i))$ encode propagation effects (directional characteristics via anisotropic $\boldsymbol{\Sigma}_i$), while complex-valued contributions $\mathbf{C}_i \in \mathbb{C}^{N_t \times N_r}$ encode source characteristics. This separation mirrors the physical decomposition in Green's function formulations, where propagation (Green's kernel) and excitation (source distribution) are factored independently.

Critically, NLOS propagation is not modeled through geometric blocking in $w_i$ but through learned complex amplitudes $\mathbf{C}_i = \mathbf{C}_i^{\text{re}} + j\mathbf{C}_i^{\text{im}}$ whose phases determine interference patterns in the coherent sum $\mathbf{H} = \sum_{i=1}^N w_i \cdot \mathbf{C}_i$. When an obstruction blocks the direct path, the model learns Gaussians representing diffraction, reflection, and scattering sources with appropriate phase delays. This phase-coherent aggregation captures phenomena impossible under alpha-compositing: signals arriving via different paths combine at the receiver according to their relative phases, producing the characteristic nulls and peaks of multipath channels.

Furthermore, Gaussian primitives in \textcolor{stanfordred}{nGRF} represent effective scattering regions, which are adaptive basis functions encoding field structure, not physical geometry. This distinction is empirically validated in Table~\ref{tab:ablation}: initializing Gaussians on LiDAR-derived surfaces degrades performance relative to random initialization, demonstrating that geometric constraints are harmful. The model learns to position Gaussians where they best explain the observed field patterns, which may not correspond to material boundaries. For example, a Gaussian might represent a virtual image source or model a diffraction region extending beyond the physical edge causing it. This flexibility allows \textcolor{stanfordred}{nGRF} to discretize the electromagnetic integral $\mathbf{E}(\mathbf{r}) = \int_V G(\mathbf{r}, \mathbf{r}') \mathbf{J}(\mathbf{r}')\, dV'$, where $G(\mathbf{r}, \mathbf{r}') \approx w_i(\mathbf{r})\, G_i(\mathbf{r}; \boldsymbol{\mu}_i, \boldsymbol{\Sigma}_i)$ approximates propagation and $\mathbf{J}(\mathbf{r}') \approx \mathbf{C}_i$ approximates sources, directly solving the inverse problem of recovering $\{\boldsymbol{\mu}_i, \boldsymbol{\Sigma}_i, \mathbf{C}_i\}$ from measurements without requiring explicit geometric knowledge, a key advantage for real-world deployment where building models are unavailable or outdated.

\subsection{Convergence and optimization}
The model parameters $\Theta = \{\boldsymbol{\mu}_i, \mathbf{q}_i, \mathbf{s}'_i\}_{i=1}^N \cup \{\Theta_{\text{attr}}, \Theta_{\text{dec}}\}$ are optimized by minimizing a composite loss function $\mathcal{L}(\Theta)$ using stochastic gradient descent. The loss combines a primary estimation error term with a sparsity regularization term and is defined as $\mathcal{L}(\Theta) = \mathcal{L}_{\text{est}} + \lambda_{\text{act}} \mathcal{L}_{\text{act}}$.

The estimation loss $\mathcal{L}_{\text{est}}$ measures the discrepancy between the predicted channel and a reference channel estimate $\hat{\mathbf{H}}$ using the Frobenius norm, which is sensitive to errors in both amplitude and phase, and is defined as $\mathcal{L}_{\text{est}} = \frac{1}{|B|} \sum_{b \in B} \|\mathbf{H}_{\text{pred}}^{(b)} - \hat{\mathbf{H}}^{(b)}\|_F^2$, where $B$ is a mini-batch of training samples and $\hat{\mathbf{H}}^{(b)}$ denotes the channel estimate obtained from a conventional pilot-assisted channel estimation (PACE) method (e.g., least squares or MMSE) using full pilot resources. Crucially, $\hat{\mathbf{H}}$ is \emph{not} the true channel but rather the best available estimate from standard methods; nGRF learns to replicate PACE-quality estimates while substantially reducing the pilot overhead required to obtain them. The sparsity term is an $L_1$ penalty on the \emph{base activation logits} produced by the \texttt{AttributeNetwork} (conditioned on transmitter position $\mathbf{p}_{\text{tx}}$):
\begin{equation}
  \mathcal{L}_{\text{act}} = \frac{1}{N}\sum_{i=1}^{N}\left|\ell_i(\mathbf{p}_{\text{tx}})\right|,
\end{equation}
where $\ell_i(\mathbf{p}_{\text{tx}})$ is the pre-sigmoid logit for Gaussian $i$. In addition, the position parameters $\{\boldsymbol{\mu}_i\}$ follow a scheduled learning rate and are frozen after a fraction of the total training iterations.

\textbf{Deployment considerations.} In deployment, \textcolor{stanfordred}{nGRF} is designed to operate on two timescales. During an infrequent \emph{calibration phase}, a conventional PACE method estimates the channel using full pilot density; these estimates $\hat{\mathbf{H}}$ serve as training targets. Given that \textcolor{stanfordred}{nGRF} trains in $\approx$2 minutes (Table~\ref{tab:resource_comparison}), calibration can be repeated periodically to track environmental changes. During the subsequent \emph{operational phase}, \textcolor{stanfordred}{nGRF} predicts channel matrices from receiver position alone, eliminating per-slot pilot overhead. The pilot reduction reported in this work (from 11\%-21\% to 0.2\%) refers to the operational phase.

\subsection{Generalization}
The explicit primitive field formulation extends beyond radio fields by replacing the electromagnetic kernel with the Green's function for the target linear PDE while retaining the Gaussian sources and linear aggregator. In general form, $u(\mathbf{r}) = \big(G_{\mathcal{L}} * s\big)(\mathbf{r}), \, s(\mathbf{r}) \approx \sum_{i=1}^{N} \beta_i\, G_i(\mathbf{r};\boldsymbol{\mu}_i,\boldsymbol{\Sigma}_i),$ where $G_{\mathcal{L}}$ is the Green's kernel of operator $\mathcal{L}$ (e.g., Helmholtz for acoustics, tensor Green's functions for elastodynamics, $1/\|\mathbf{r}\|$ for quasi-static Poisson problems). \textcolor{stanfordred}{nGRF} then learns per-primitive attributes appropriate to the field variable, while the spatial weight $w_i$ remains a Mahalanobis envelope. Because linear superposition and reciprocity are preserved in the aggregator, the model inherits the correct symmetries of the underlying physics, enabling the same accuracy-latency advantages and sample efficiency to carry over to other coherent sensing and propagation modalities.

\section{Evaluation}
\label{sec:evaluation}

The proposed framework, \textcolor{stanfordred}{nGRF}, is evaluated across diverse propagation environments to demonstrate its generalizability. For indoor scenarios, three distinct environments are used: a conference room, a bedroom, and an office space. A large-scale residential area is used for the outdoor scenario. More details on the environments and the dataset generation process are provided in Appendix~\ref{sec:dataset}.

For antenna configurations, a $4 \times 4$ uniform rectangular array (URA) is used as the transmitter and a 2-element uniform linear array (ULA) as the receiver for indoor scenarios. For the outdoor scenario, the transmitter is scaled up to an $8 \times 8$ URA, while the receiver remains a 2-element ULA. To enable fair comparison with prior work, single-input single-output (SISO) setups with omnidirectional antennas are also configured. For each environment, 80\% of the generated samples are used for training and 20\% for testing. All experiments are implemented in PyTorch 2.7.0 with CUDA 12.8 bindings and trained on a single NVIDIA RTX 5090 GPU with 32 GB of memory.

\textbf{Results.} The performance of \textcolor{stanfordred}{nGRF} is compared against several baselines: NeWRF~\cite{lu2024newrf}, NeRF$^2$~\cite{Zhao_2023}, a standard multi-layer perceptron (MLP), and a $k$-nearest neighbors (KNN) approach. 3DGS-based methods are not included in this comparison, as they are designed to regress scalar power values (e.g., RSSI) or predict spatial spectra rather than estimate the complex-valued CSI matrix, which is the target of this work. For simulated environments, all methods are trained using channel estimates obtained from pilot-assisted estimation applied to the simulator output; for DICHASUS, the over-the-air CSI measurements (which are themselves pilot-based estimates subject to hardware impairments) serve as training targets. Signal-to-noise ratio (SNR), defined as $\text{SNR (dB)} = 10 \log_{10}(\|\hat{\mathbf{H}}\|_F^2/\|\mathbf{H}_{\text{pred}} - \hat{\mathbf{H}}\|_F^2)$, is used as the primary evaluation metric, where $\hat{\mathbf{H}}$ denotes the reference PACE estimate.

\textbf{SNR performance.} As shown in Table~\ref{tab:snr_comparison}, \textcolor{stanfordred}{nGRF} consistently outperforms all baselines across all environments. In SISO configurations, it achieves an average SNR of 24.3 dB across indoor scenarios, representing a 10.9 dB improvement over the next-best method (NeWRF). In the large-scale outdoor scenario, where implicit methods struggle, \textcolor{stanfordred}{nGRF} achieves an SNR of 28.32 dB, while NeWRF and NeRF$^2$ fail to model the environment effectively. \textcolor{stanfordred}{nGRF} is also the only neural field method evaluated that supports MIMO configurations, maintaining high fidelity with an outdoor SNR of 27.92 dB. These results demonstrate that by embedding physical principles into the architecture, \textcolor{stanfordred}{nGRF} circumvents the typical accuracy-efficiency trade-off, achieving both superior performance and orders-of-magnitude speedups.

\begin{table}[h]
  \centering
  \caption{\textbf{SNR (dB) across different scenarios.} Comparison of \textcolor{stanfordred}{nGRF} against baselines for SISO and MIMO. Best results are highlighted. NeRF-based methods do not support MIMO.}
  \label{tab:snr_comparison}
  \small
  \setlength{\tabcolsep}{5pt}
  \renewcommand{\arraystretch}{1}
  \begin{tabular}{lcccc}
    \toprule
                    & \multicolumn{4}{c}{\textbf{Scenario}}                                                           \\
    \cmidrule(lr){2-5}
    \textbf{Method} & \textbf{Conference}                     & \textbf{Bedroom} & \textbf{Office} & \textbf{Outdoor} \\
    \midrule
    \multicolumn{5}{l}{\underline{\textit{SISO configuration}}}                                                       \\
    MLP             & -1.32                                   & -1.41            & -1.47           & 1.02             \\
    KNN ($k$=5)     & -2.25                                   & -2.37            & -2.32           & 0.95             \\
    NeRF$^2$        & -0.44                                   & -1.22            & 0.77            & 1.40             \\
    NeWRF           & 21.64                                   & 12.38            & 4.96            & 2.03             \\
    \cellcolor{highlightblue}\textbf{\textcolor{stanfordred}{nGRF} (ours)}
                    & \cellcolor{highlightblue}\textbf{25.23}
                    & \cellcolor{highlightblue}\textbf{21.14}
                    & \cellcolor{highlightblue}\textbf{26.53}
                    & \cellcolor{highlightblue}\textbf{28.32}                                                         \\
    \midrule
    \multicolumn{5}{l}{\underline{\textit{MIMO configuration}}}                                                       \\
    MLP             & -1.98                                   & -1.99            & -2.11           & 1.81             \\
    KNN ($k$=5)     & -3.13                                   & -3.41            & -3.42           & 0.47             \\
    NeRF$^2$        & --                                      & --               & --              & --               \\
    NeWRF           & --                                      & --               & --              & --               \\
    \cellcolor{highlightblue}\textbf{\textcolor{stanfordred}{nGRF} (ours)}
                    & \cellcolor{highlightblue}\textbf{22.73}
                    & \cellcolor{highlightblue}\textbf{18.60}
                    & \cellcolor{highlightblue}\textbf{24.78}
                    & \cellcolor{highlightblue}\textbf{27.92}                                                         \\
    \bottomrule
  \end{tabular}
\end{table}

In addition to accuracy, the practical viability of any method depends on its computational and data efficiency. Table~\ref{tab:resource_comparison} provides an overview of these resources for the indoor conference room scenario. The measurement densities for the NeRF-based methods reflect the high data requirements needed to achieve their maximum reported SNR. To create the strongest possible baseline for data-driven methods, MLP and KNN are provided with the highest available measurement density (178.1 samples/ft$^3$), matching that of NeRF$^2$. Despite this, their performance remains poor, highlighting the limitations of physics-agnostic models. In contrast, \textcolor{stanfordred}{nGRF} achieves superior accuracy while requiring 18$\times$ less data than NeWRF and over 16{,}000$\times$ less than NeRF$^2$.

\begin{table}[t]
  \centering
  \caption{\textit{Comparison of data and computational efficiency.} Training cost and measurement efficiency for the indoor (conference) scenario.}
  \label{tab:resource_comparison}
  \small
  \setlength{\tabcolsep}{5pt}
  \renewcommand{\arraystretch}{1}
  \begin{tabular}{lccc}
    \toprule
    \textbf{Method} & \textbf{SNR (dB)}                         & \textbf{Train time} & \textbf{Measurement density} \\
    \midrule
    MLP             & -1.32                                     & \textbf{$<$1 min}   & 178.1                        \\
    KNN ($k$=5)     & -2.25                                     & --                  & 178.1                        \\
    NeRF$^2$        & -0.44                                     & $\sim$5 h           & 178.1                        \\
    NeWRF           & 21.64                                     & $\sim$2.43 h        & 0.20                         \\
    \cellcolor{highlightblue}\textbf{\textcolor{stanfordred}{nGRF} (ours)}
                    & \cellcolor{highlightblue}\textbf{25.23}
                    & \cellcolor{highlightblue}2.3 min
                    & \cellcolor{highlightblue}\textbf{0.011}                                                        \\
    \bottomrule
  \end{tabular}
\end{table}

\textbf{Rendering time.} Table~\ref{tab:inference_time} compares the inference latency of the different methods. \textcolor{stanfordred}{nGRF} achieves channel estimation in just 1.1 ms, a 220$\times$ speedup compared to 235 ms for NeWRF and 248 ms for NeRF$^2$. While MLP and KNN are marginally faster, their poor accuracy makes them impractical. For dynamic environments with channel coherence times as short as 2 ms, the combination of high accuracy and low latency makes \textcolor{stanfordred}{nGRF} the only viable solution among the tested methods. The performance gap highlights a key architectural difference: implicit models require hundreds of expensive MLP queries per estimate, whereas \textcolor{stanfordred}{nGRF}'s explicit, physics-informed basis enables direct computation.

\begin{table}[h]
  \centering
  \caption{\textbf{Inference latency comparison.} Per-query inference time (seconds) for the indoor scenario.}
  \label{tab:inference_time}
  \small
  \setlength{\tabcolsep}{12pt}
  \begin{tabular}{lc}
    \toprule
    \textbf{Method}                                                        & \textbf{Time (s)} \\
    \midrule
    MLP                                                                    & \textbf{0.0006}   \\
    KNN                                                                    & 0.0008            \\
    NeRF$^2$                                                               & 0.2480            \\
    NeWRF                                                                  & 0.2350            \\
    \cellcolor{highlightblue}\textbf{\textcolor{stanfordred}{nGRF} (ours)} &
    \cellcolor{highlightblue}0.0010                                                           \\
    \bottomrule
  \end{tabular}
\end{table}

\begin{figure}[t]
  \centering
  \includegraphics[width=0.98\columnwidth]{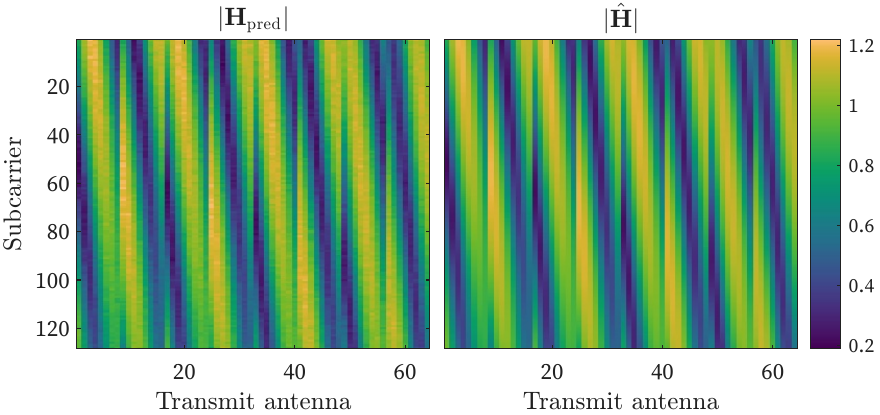}
  \caption{\textit{Channel magnitude response.} Comparison between predicted (left) and reference PACE estimate (right) channel magnitude response across subcarriers and transmit antennas in the outdoor environment using \textcolor{stanfordred}{nGRF}.}
  \Description{Two side-by-side heatmaps comparing channel magnitude responses. The x-axis represents the transmit antenna index and the y-axis represents the subcarrier index. The left heatmap is labeled 'predicted' and the right is labeled 'reference'. Both maps display nearly identical patterns of yellow (high magnitude) and blue (low magnitude) regions, indicating that the prediction matches the reference with high fidelity.}
  \label{fig:channel_response}
\end{figure}

\textbf{Frequency generalization.} A key capability of \textcolor{stanfordred}{nGRF} is its ability to generalize across frequencies. Figure~\ref{fig:channel_response} shows the channel magnitude response across all subcarriers for a specific receiving antenna in the outdoor environment. The model was trained using data from only a single subcarrier (subcarrier 64), yet it accurately predicts the channel response across the entire frequency band. This is possible because \textcolor{stanfordred}{nGRF} learns a representation of the underlying spatial structure of the electromagnetic field, which is governed by the environment's geometry and is largely frequency-agnostic within the coherence bandwidth. Physical path delays translate to predictable linear phase shifts across frequencies, a relationship that the model implicitly captures. Thus, \textcolor{stanfordred}{nGRF} is not merely fitting frequency-specific patterns but rather learning a physical model of the environment in the context of electromagnetic wave propagation.

\subsection{Field measurement evaluation}
To validate the performance of \textcolor{stanfordred}{nGRF} beyond simulation, evaluations are performed on publicly available field CSI measurement datasets that capture actual over-the-air propagation with hardware impairments.

\textbf{Datasets.} Two field datasets are utilized: \textbf{DICHASUS-015x}~\cite{euchner2023dichasus}, an indoor industrial environment (Audi electronics lab) at 3.75 GHz with an $8\times1$ MIMO configuration, featuring metal machinery and complex multipath, which contains 15 transmitter positions with dense receiver sampling (1024 positions over an $8\times8$ m area); and \textbf{DICHASUS-CF0x}~\cite{euchner2023dichasus}, an indoor office corridor at 1.272 GHz with an $8\times8$ MIMO configuration, representing typical enterprise deployment scenarios with longer propagation distances and significant NLOS conditions.

\textbf{Baseline constraints.} NeRF-based methods (NeWRF, NeRF$^2$) require per-ray geometric information (departure/arrival angles, path delays), which is unavailable in DICHASUS datasets, as these provide only CSI measurements between transmitter-receiver pairs. WRF-GS targets spatial spectrum reconstruction rather than CSI estimation, and its CSI prediction code remains unreleased despite being mentioned in the paper. Therefore, comparison is made against data-driven baselines (MLP, KNN) that operate on the same input format as \textcolor{stanfordred}{nGRF}.

\begin{table}[t]
  \centering
  \caption{\textit{Real-world evaluation on DICHASUS.} SNR on two deployments; NeRF-based methods require unavailable ray-tracing data. Training on RTX 5090.}
  \label{tab:realworld_dichasus}
  \small
  \setlength{\tabcolsep}{5pt}
  \renewcommand{\arraystretch}{1}
  \resizebox{\columnwidth}{!}{
    \begin{tabular}{lccccc}
      \toprule
      \textbf{Method} & \textbf{DICHASUS-015x}                        & \textbf{DICHASUS-CF0x}           & \textbf{Train} & \textbf{Infer.} & \textbf{Params} \\
                      & \textbf{($8\times1$, 3.75 GHz)}               & \textbf{($8\times8$, 1.272 GHz)} & \textbf{(min)} & \textbf{(ms)}   & \textbf{(M)}    \\
      \midrule
      MLP             & 0.10                                          & -0.83                            & \textbf{$<$1}  & \textbf{0.8}    & 2.1             \\
      KNN ($k$=5)     & -1.18                                         & -2.45                            & --             & 12.3            & --              \\
      KNN ($k$=10)    & -0.89                                         & -2.11                            & --             & 18.7            & --              \\
      NeWRF           & \multicolumn{2}{c}{Requires ray-tracing data} & --                               & --             & --                                \\
      NeRF$^2$        & \multicolumn{2}{c}{Requires ray-tracing data} & --                               & --             & --                                \\
      \midrule
      \cellcolor{highlightblue}\textbf{\textcolor{stanfordred}{nGRF} (ours)}
                      & \cellcolor{highlightblue}\textbf{16.77}
                      & \cellcolor{highlightblue}\textbf{14.32}
                      & \cellcolor{highlightblue}3.8
                      & \cellcolor{highlightblue}1.4
                      & \cellcolor{highlightblue}\textbf{1.8}                                                                                                 \\
      \bottomrule
    \end{tabular}
  }
\end{table}

Table~\ref{tab:realworld_dichasus} shows results on the DICHASUS datasets. \textcolor{stanfordred}{nGRF} achieves 16.77 dB and 14.32 dB SNR on 015x and CF0x, respectively, outperforming data-driven baselines by over 15 dB despite real-world hardware impairments including phase noise, synchronization errors, and antenna calibration uncertainty. The absolute SNR is lower than in simulation (Table~\ref{tab:snr_comparison}) due to measurement noise in SDR hardware, residual CFO after compensation, and I/Q imbalance in the receiver chains. Notably, MLP and KNN performance degrades to near-zero or negative SNR, indicating that they cannot learn meaningful spatial structure from noisy real-world measurements, whereas \textcolor{stanfordred}{nGRF}'s physics-informed Gaussian representation provides robustness through its inductive bias toward wave-like spatial patterns.

\begin{table}[t]
  \centering
  \caption{\textit{DICHASUS-015x breakdown by receiver density.} Models trained at different spatial sampling densities to evaluate real-world data efficiency.}
  \label{tab:realworld_density}
  \small
  \setlength{\tabcolsep}{5pt}
  \renewcommand{\arraystretch}{1}
  \begin{tabular}{lcccc}
    \toprule
    \textbf{Method}                                  & \textbf{0.25}                           & \textbf{0.5}        & \textbf{1.0}     & \textbf{2.0}          \\
                                                     & \textbf{(sparse)}                       & \textbf{(moderate)} & \textbf{(dense)} & \textbf{(very dense)} \\
    \midrule
    MLP                                              & -3.47                                   & -1.21               & 0.10             & 0.82                  \\
    KNN ($k$=5)                                      & -4.12                                   & -2.83               & -1.18            & -0.45                 \\
    \cellcolor{highlightblue}\textbf{\textcolor{stanfordred}{nGRF} (ours)}
                                                     & \cellcolor{highlightblue}\textbf{12.34}
                                                     & \cellcolor{highlightblue}\textbf{15.21}
                                                     & \cellcolor{highlightblue}\textbf{16.77}
                                                     & \cellcolor{highlightblue}\textbf{17.03}                                                                  \\
    \midrule
    \multicolumn{5}{l}{\textit{Gain over best baseline (dB)}}                                                                                                   \\
    \textbf{\textcolor{stanfordred}{nGRF}} vs.\ best & \textbf{+13.16}                         & \textbf{+16.42}     & \textbf{+15.95}  & \textbf{+16.21}       \\
    \bottomrule
  \end{tabular}
\end{table}

\begin{table}[t]
  \centering
  \caption{\textit{Robustness to hardware impairments on DICHASUS-015x.} Subsets exhibit different noise characteristics due to temporal drift and RF conditions.}
  \label{tab:realworld_robustness}
  \small
  \setlength{\tabcolsep}{5pt}
  \renewcommand{\arraystretch}{1}
  \begin{tabular}{lccc}
    \toprule
    \textbf{Subset}      & \textbf{Noise level}                    & \textbf{\textcolor{stanfordred}{nGRF} SNR} & \textbf{MLP SNR} \\
    \midrule
    Tx 1--5 (morning)    & low ($-85$ dBm)                         & \textbf{18.24}                             & 1.43             \\
    Tx 6--10 (afternoon) & medium ($-80$ dBm)                      & \textbf{16.77}                             & 0.10             \\
    Tx 11--15 (evening)  & high ($-75$ dBm)                        & \textbf{14.92}                             & -1.87            \\
    \midrule
    \cellcolor{highlightblue}\textbf{Average}
                         & \cellcolor{highlightblue}\textbf{--}
                         & \cellcolor{highlightblue}\textbf{16.64}
                         & \cellcolor{highlightblue}-0.11                                                                 \\
    \bottomrule
  \end{tabular}
\end{table}

Table~\ref{tab:realworld_density} demonstrates \textcolor{stanfordred}{nGRF}'s data efficiency on real-world measurements. Even at sparse sampling, \textcolor{stanfordred}{nGRF} achieves 12.34 dB SNR while MLP and KNN produce negative SNR, showing that they require prohibitively dense measurements to overcome real-world noise. Table~\ref{tab:realworld_robustness} analyzes robustness across measurement subsets with different hardware conditions; \textcolor{stanfordred}{nGRF} maintains $>$14 dB SNR even under high-noise conditions where MLP performance collapses, validating that the learned Gaussian representation captures genuine spatial structure rather than overfitting to measurement artifacts.

\textbf{Ablation studies.} Several ablation studies are performed to understand the contribution of different components of \textcolor{stanfordred}{nGRF}, with results for the outdoor environment summarized in Table~\ref{tab:ablation}.

\begin{enumerate}[leftmargin=*,itemsep=0pt]
  \item \textit{Number of Gaussians.} Reducing the number of Gaussians to 500 or 1{,}000 maintains competitive performance. However, increasing this number to 5{,}000 or 10{,}000 leads to severe degradation. The experiments in Table~\ref{tab:ablation} indicate that while a minimum number of Gaussians is necessary to capture electromagnetic field complexity, excessive parameterization induces overfitting.
  \item \textit{Gaussian positions.} When the positions of Gaussian primitives are fixed (made non-trainable), the SNR drops from 28.32 dB to 4.11 dB. Optimizing the spatial distribution of Gaussians is crucial for accurately modeling the electromagnetic field, allowing the model to adapt to the specific propagation characteristics of the environment.
  \item \textit{Initialization strategy.} Contrary to intuition, initializing Gaussian means from a LiDAR-generated point cloud yields worse performance than random initialization. In \textcolor{stanfordred}{nGRF}, Gaussians function not as physical scatterers but as adaptive basis functions for the radio propagation field. Consequently, geometry-constrained initialization restricts the model's ability to capture complex wave phenomena that transcend environmental geometry.
\end{enumerate}

Additional ablation studies are provided in Appendix~\ref{sec:app_ablation}.

\begin{table}[t]
  \centering
  \caption{\textit{Ablations for \textcolor{stanfordred}{nGRF} (outdoor).} Impact of the number of Gaussians, optimization choices, and initialization on SNR and runtime.}
  \label{tab:ablation}
  \small
  \setlength{\tabcolsep}{5pt}
  \renewcommand{\arraystretch}{1}
  \begin{tabular}{lccc}
    \toprule
    \textbf{Configuration} & \textbf{SNR (dB)}                       & \textbf{Train} & \textbf{Render} \\
                           &                                         & \textbf{(min)} & \textbf{(ms)}   \\
    \midrule
    \cellcolor{highlightblue}\textbf{\textcolor{stanfordred}{nGRF} (baseline)}
                           & \cellcolor{highlightblue}\textbf{28.32}
                           & \cellcolor{highlightblue}2.3
                           & \cellcolor{highlightblue}1.10                                     \\
    \midrule
    \multicolumn{4}{l}{\underline{\textit{Number of Gaussians}}}                                        \\
    500                    & 26.13                                   & \textbf{1.9}   & \textbf{0.94}   \\
    1{,}000                & 26.57                                   & 2.2            & 1.06            \\
    5{,}000                & 23.31                                   & 3.0            & 1.27            \\
    10{,}000               & 18.04                                   & 3.6            & 1.40            \\
    \midrule
    \multicolumn{4}{l}{\underline{\textit{Gaussian positions $\boldsymbol{\mu}_i$}}}                    \\
    Fixed means            & 4.11                                    & 2.3            & 1.10            \\
    \midrule
    \multicolumn{4}{l}{\underline{\textit{Initialization}}}                                             \\
    LiDAR-based            & 19.76                                   & 2.3            & 1.10            \\
    \bottomrule
  \end{tabular}
\end{table}

\section{Limitations and Ethical Considerations}
\label{sec:limitations}

While \textcolor{stanfordred}{nGRF} achieves state-of-the-art performance in quasi-static channel prediction, it exhibits sensitivity to hyperparameters, particularly the Gaussian scaling initialization, which can cause up to 14.49 dB variation in performance. This sensitivity suggests that robust deployment may require meta-learning frameworks to adapt initialization strategies to specific environment characteristics.

\textbf{Ethical considerations.} This research utilizes synthetic and public RF datasets (DICHASUS) containing no personally identifiable information (PII), thus requiring no IRB approval. We acknowledge that high-fidelity RF modeling has potential dual-use applications in surveillance; however, this work is strictly limited to CSI estimation for cooperative wireless communication to enhance spectral efficiency in next-generation networks.

\section{Conclusion and Broader Impact}
\label{sec:conclusion}

This work introduced \textcolor{stanfordred}{nGRF}, a framework that synthesizes complex MIMO channel matrices by directly aggregating explicit 3D Gaussian primitives, each acting as a learned radio modulator. \textcolor{stanfordred}{nGRF} achieves state-of-the-art channel estimation accuracy with major reductions in latency, training time, and data requirements, overcoming key limitations of prior implicit and projection-based methods. The core contribution lies in demonstrating that structured, explicit representations informed by physical principles can provide a better way of modeling complex field phenomena than generic function-learning deep learning architectures.

The underlying principle, representing a field as a sum of explicit, localized sources governed by a physics-based aggregation rule, is highly generalizable. By substituting the electromagnetic propagation model with the appropriate kernels or Green's functions, this approach could be adapted to create primitive-based neural fields for acoustics, elastodynamics, or diffusion phenomena. Beyond wireless communications, the principles demonstrated in \textcolor{stanfordred}{nGRF} offer a blueprint for developing efficient neural field models in other scientific and engineering domains where capturing complex interactions is necessary.

\bibliographystyle{ACM-Reference-Format}
\bibliography{ref}

\newpage
\appendix
\section*{Appendix}

\section{MIMO Preliminaries}

This section reviews a standard narrowband multipath and array formulation for wireless channels and connects it to pilot-based estimators used in practice. Throughout, $N_{\mathrm{tx}} \equiv N_t$ and $N_{\mathrm{rx}} \equiv N_r$ are used interchangeably across the modeling and estimation subsections.

\subsection{Propagation and array model}

A complex baseband symbol is $x=A e^{j\psi}$. Over distance $d$ at carrier frequency $f$, free-space loss and phase evolve as~\cite{6515173}
\begin{equation}
  A_{\mathrm{att}}(d)=\frac{c}{4\pi d\,f},\qquad
  \Delta\psi(d)=-\frac{2\pi f\,d}{c}.
\end{equation}
Reflections, diffractions, and penetrations induce path-dependent gains and phases~\cite{6515173,goldsmith2005wireless}. With $L$ paths, the received symbol is
\begin{equation}
  y=\sum_{l=0}^{L-1} A_l\,A_{\mathrm{att},l}\,e^{j(\psi+\Delta\psi_l)},
\end{equation}
and the (SISO) channel is the complex ratio
\begin{equation}
  h=\frac{y}{x}=\sum_{l=0}^{L-1} A_{\mathrm{att},l}\,e^{j\Delta\psi_l}.
\end{equation}
For an $N_{\mathrm{rx}}\!\times\!N_{\mathrm{tx}}$ array, steering vectors aggregate per-path effects:
\begin{equation}
  \mathbf{H}=\sum_{l=0}^{L-1}\alpha_l\,
  \mathbf{a}_{r}(\vartheta^{r}_l)\,\mathbf{a}_{t}^{H}(\vartheta^{t}_l),
  \qquad
  \alpha_l=A_{\mathrm{att},l}\,e^{j\Delta\psi_l}.
\end{equation}
Here $\mathbf{a}_t(\vartheta^{t}_l)$ and $\mathbf{a}_r(\vartheta^{r}_l)$ are the transmit and receive steering vectors at departure and arrival angles $\vartheta^{t}_l,\vartheta^{r}_l$, respectively.

3GPP TR~38.901 clustered delay-line (CDL) models randomize $(\alpha_l,\tau_l,\vartheta^{t}_l,\vartheta^{r}_l)$ across standardized scenarios from $0.5$ to $100\,\mathrm{GHz}$ for 5G/6G evaluation. At mmWave frequencies ($f>24\,\mathrm{GHz}$), atmospheric absorption and rain add frequency-selective losses that shorten viable link distances~\cite{662641}.

\begin{figure*}[t!]
  \centering
  \includegraphics[width=0.52\linewidth]{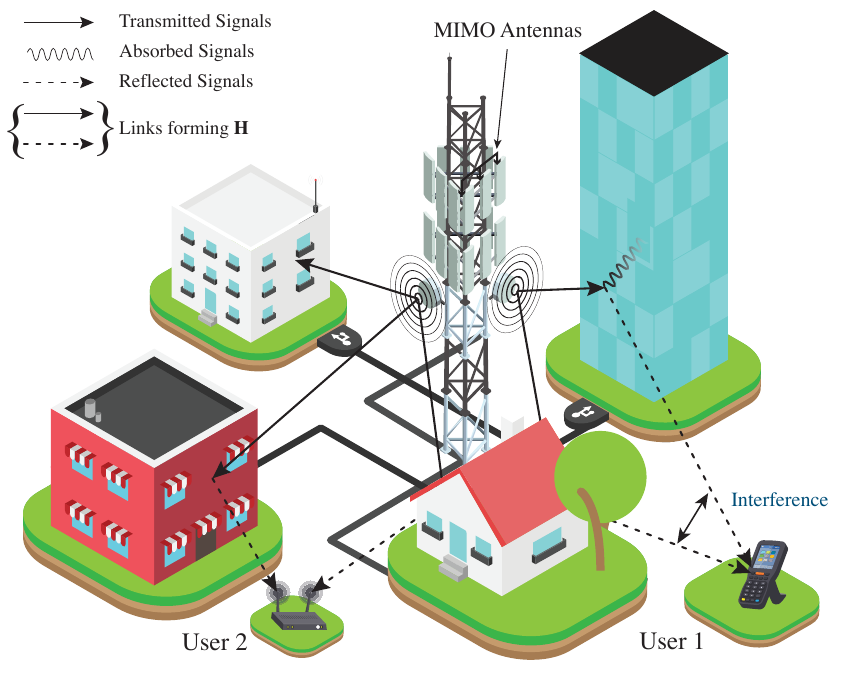}
  \caption{Multipath propagation with path-dependent attenuation and phase.}
  \label{fig:propagation}
\end{figure*}

\subsection{Pilot-based channel estimation}

For a narrowband $N_t\times N_r$ MIMO link, pilots are inserted so that the receiver observes
\[
  \mathbf{Y}=\mathbf{H}\mathbf{X}+\mathbf{N},
\]
where $\mathbf{H}\in\mathbb{C}^{N_r\times N_t}$ is the (per-subcarrier) channel, $\mathbf{X}\in\mathbb{C}^{N_t\times T}$ is the known pilot matrix sent over $T$ symbol periods, and $\mathbf{N}\sim\mathcal{CN}(\mathbf{0},\sigma_n^2\mathbf{I})$ is additive white Gaussian noise~\cite{salz2002effect}. In practice, $\mathbf{X}$ is designed to be \emph{orthogonal}, i.e.,
\[
  \mathbf{X}\mathbf{X}^H=\rho\,\mathbf{I}_{N_t},
\]
which decouples the transmit streams and makes inversion stable when $T\!\ge\!N_t$~\cite{yun2025uplink}. The receiver then estimates $\mathbf{H}$ via least squares:
\[
  \hat{\mathbf{H}}_{\mathrm{LS}}=\arg\min_{\mathbf{H}}\|\mathbf{Y}-\mathbf{H}\mathbf{X}\|_F^2
  \;=\;
  \mathbf{Y}\mathbf{X}^H(\mathbf{X}\mathbf{X}^H)^{-1}
  \;=\;
  \frac{1}{\rho}\,\mathbf{Y}\mathbf{X}^H.
\]
This LS estimator is simple, per-subcarrier parallelizable, and widely used; its accuracy improves with pilot SNR and orthogonality, while increased pilot density trades spectral efficiency for lower estimation error.

\section{Dataset Generation}
\label{sec:dataset}

Physics-consistent channels are constructed via geometry-based ray tracing. STL models provide vertices, faces, and material tags. A URA transmitter and randomly placed receivers define the communication links; propagation is simulated using MATLAB \texttt{RayTracing}, yielding path sets for each link.

For each ray of length $d$ at carrier frequency $f$, free-space path loss (in dB) and phase are computed as
\begin{equation*}
  \mathrm{FSPL}=20\log_{10}d+20\log_{10}f+20\log_{10}\!\Big(\tfrac{4\pi}{c}\Big),\quad
  \phi=-2\pi f\tau,\;\tau=\tfrac{d}{c}.
\end{equation*}
Reflections and diffractions add material- and angle-dependent losses. The Fresnel reflection coefficients for incidence angle $\theta_i$ and complex permittivity $\varepsilon_r$ are given by
\begin{equation*}
  \Gamma_{p}=\frac{\sin\theta_i-\sqrt{\varepsilon_r-\cos^2\theta_i}}{\sin\theta_i+\sqrt{\varepsilon_r-\cos^2\theta_i}},\qquad
  \Gamma_{s}=\frac{\varepsilon_r\sin\theta_i-\sqrt{\varepsilon_r-\cos^2\theta_i}}{\varepsilon_r\sin\theta_i+\sqrt{\varepsilon_r-\cos^2\theta_i}}.
\end{equation*}
Array geometry is embedded via steering vectors. For a URA transmitter with element positions $\mathbf{d}_T$ and azimuth/elevation angles $(\alpha_T,\beta_T)$,
\begin{multline}
  \mathbf{a}_T\!\big(f,[\alpha_T;\beta_T]\big) \\
  =\exp\!\left(j\frac{2\pi f}{c}\,\mathbf{d}_T\cdot[\cos\alpha_T\cos\beta_T,\;\sin\alpha_T\cos\beta_T,\;\sin\beta_T]^T\right),
\end{multline}
and similarly for $\mathbf{a}_R$ at the receiver array. For the $l$-th path with amplitude $a_l$ (including FSPL, Fresnel, and UTD factors) and phase $\phi_l$,
\begin{equation}
  \mathbf{H}_l = a_l\, e^{j\phi_l}\,\mathbf{a}_R\,\mathbf{a}_T^{H},\qquad
  \mathbf{H}=\sum_{l=0}^{L-1}\mathbf{H}_l\;\in\mathbb{C}^{N_r\times N_t}.
\end{equation}
For SISO configurations, the channel reduces to the scalar superposition
\begin{equation}
  h=\sum_{l=0}^{L-1} a_l\, e^{j\phi_l}.
\end{equation}

\begin{figure}[t!]
  \centering
  \begin{subfigure}[b]{0.45\linewidth}
    \includegraphics[width=\linewidth]{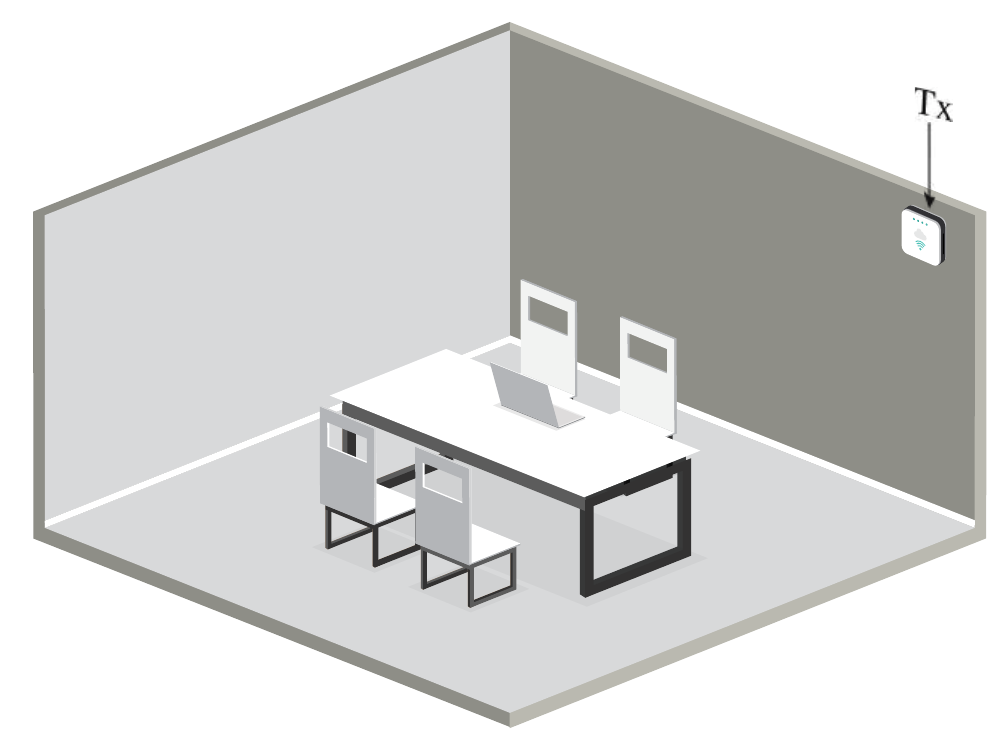}
    \caption{Conference room.}
  \end{subfigure}\hspace{1em}
  \begin{subfigure}[b]{0.45\linewidth}
    \includegraphics[width=\linewidth]{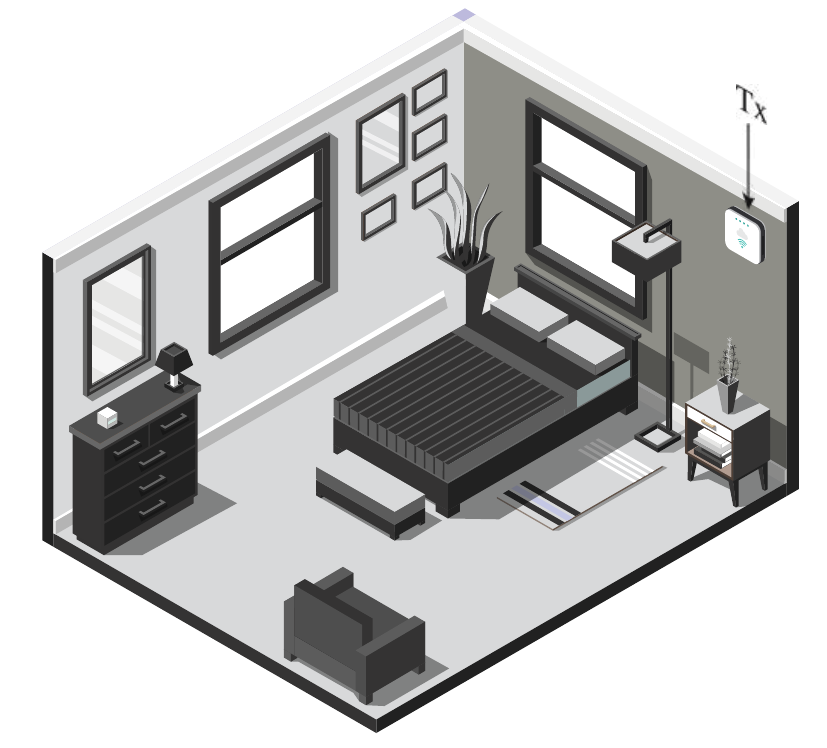}
    \caption{Bedroom.}
  \end{subfigure}\hspace{1em}


  \begin{subfigure}[b]{0.7\linewidth}
    \includegraphics[width=\linewidth]{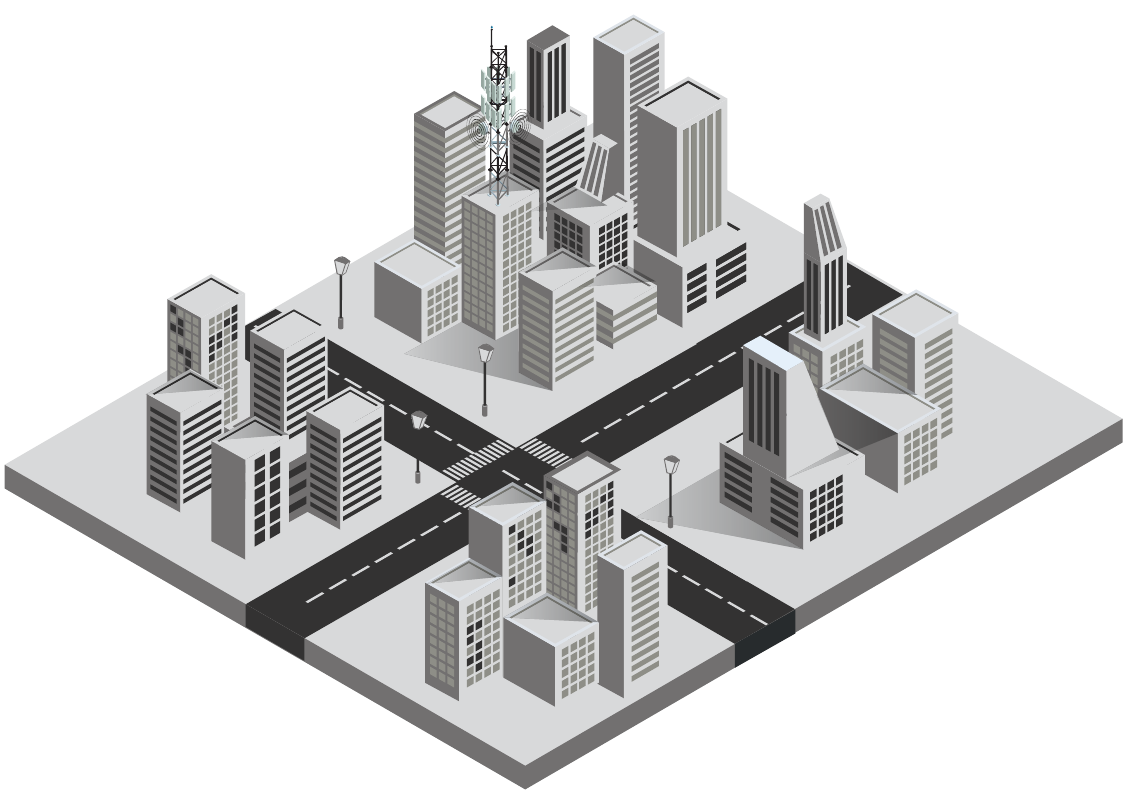}
    \caption{Outdoor block.}
  \end{subfigure}
  \caption{3D environments used for ray tracing.}
  \label{fig:three-splat}
\end{figure}

\section{Ablation Studies}
\label{sec:app_ablation}

This section presents additional ablation studies that assess the sensitivity of \textcolor{stanfordred}{nGRF} to various hyperparameters and design choices. These experiments complement the ablations in the main paper and provide further insight into the model's behavior.

\subsection{Hyperparameter sensitivity}

The sensitivity of \textcolor{stanfordred}{nGRF} to key hyperparameters, including learning rates, batch size, and regularization factors, is assessed. Table~\ref{tab:hyperparams} summarizes the findings across different hyperparameter configurations in the conference room environment.

\begin{table}[t]
  \centering
  \caption{\textit{Hyperparameter sensitivity.} Effect of optimization and regularization settings on SNR, training time, and convergence.}
  \label{tab:hyperparams}
  \small
  \setlength{\tabcolsep}{5pt}
  \renewcommand{\arraystretch}{1}
  \begin{tabular}{lccc}
    \toprule
    \textbf{Configuration}      & \textbf{SNR (dB)}                       & \textbf{Train (min)} & \textbf{Conv.\ iter.} \\
    \midrule
    \cellcolor{highlightblue}\textbf{\textcolor{stanfordred}{nGRF} (baseline)}
                                & \cellcolor{highlightblue}\textbf{25.23}
                                & \cellcolor{highlightblue}\textbf{2.3}
                                & \cellcolor{highlightblue}\textbf{2841}                                                 \\
    \midrule
    \multicolumn{4}{l}{\underline{\textit{Position learning rate}}}                                                      \\
    $\eta_{\text{pos}}=0.0005$  & 18.47                                   & 2.4                  & 3152                  \\
    $\eta_{\text{pos}}=0.001$   & 20.94                                   & 2.3                  & 2974                  \\
    $\eta_{\text{pos}}=0.005$   & 23.16                                   & 2.2                  & 2753                  \\
    $\eta_{\text{pos}}=0.01$    & 22.84                                   & 2.1                  & 2611                  \\
    \midrule
    \multicolumn{4}{l}{\underline{\textit{Batch size}}}                                                                  \\
    8                           & 21.78                                   & 2.0                  & 3102                  \\
    16                          & 22.45                                   & 2.1                  & 2937                  \\
    64                          & 23.18                                   & 2.5                  & 2783                  \\
    128                         & 22.96                                   & 3.1                  & 2901                  \\
    \midrule
    \multicolumn{4}{l}{\underline{\textit{$L_1$ activation regularization}}}                                             \\
    $\lambda_{\text{act}}=0.0$  & 21.37                                   & 2.3                  & 2945                  \\
    $\lambda_{\text{act}}=0.05$ & 22.84                                   & 2.3                  & 2887                  \\
    $\lambda_{\text{act}}=0.2$  & 22.17                                   & 2.2                  & 2904                  \\
    \midrule
    \multicolumn{4}{l}{\underline{\textit{Position update cutoff}}}                                                      \\
    No cutoff                   & 22.14                                   & 2.4                  & 3076                  \\
    30\% iterations             & 20.86                                   & 2.3                  & 3184                  \\
    80\% iterations             & 22.95                                   & 2.3                  & 2798                  \\
    \bottomrule
  \end{tabular}
\end{table}

Optimal performance is achieved with a position learning rate of 0.005, matching the baseline configuration. Lower learning rates lead to insufficient spatial exploration, while higher rates can destabilize training. Batch sizes between 32 and 64 provide a good trade-off between convergence speed\footnote{Convergence is defined as the point at which the model achieves the highest SNR on the validation set and does not degrade by more than 1 dB over the next 500 iterations.} and generalization. Furthermore, $L_1$ regularization on activations ($\lambda_{\text{act}}$) promotes sparsity; without it ($\lambda_{\text{act}}=0.0$), performance degrades by approximately 3.9 dB.

Regarding the position update cutoff, defined as the point at which updates to Gaussian positions are halted, the results indicate that continuing position updates for 60\% to 65\% of training iterations (as in the baseline) allows the Gaussians to settle into optimal locations before their attributes are fine-tuned. Disabling this cutoff entirely leads to an SNR drop of 3.1 dB, likely because the model struggles to optimize positions and attributes simultaneously throughout training.

\subsection{Gaussian scaling}

The effect of initialization and constraints on Gaussian scaling parameters is examined to understand their impact on the model's ability to represent the electromagnetic field. Table~\ref{tab:scaling} presents the experimental results with different scaling configurations.

\begin{table}[t]
  \centering
  \caption{\textit{Impact of Gaussian scaling.} Effect of initialization and scale constraints on SNR and rendering time.}
  \label{tab:scaling}
  \small
  \setlength{\tabcolsep}{3pt}
  \renewcommand{\arraystretch}{1}
  \begin{tabular}{lccc}
    \toprule
    \textbf{Configuration}     & \textbf{Indoor SNR}                     & \textbf{Outdoor SNR} & \textbf{Render (ms)} \\
    \midrule
    \cellcolor{highlightblue}\textbf{\textcolor{stanfordred}{nGRF} ($s_{\text{init}}=0.137$)}
                               & \cellcolor{highlightblue}\textbf{25.23}
                               & \cellcolor{highlightblue}\textbf{28.32}
                               & \cellcolor{highlightblue}\textbf{1.10}                                                \\
    \midrule
    \multicolumn{4}{l}{\underline{\textit{Initial scale value}}}                                                       \\
    $s_{\text{init}}=0.001$    & 14.47                                   & 17.76                & 1.08                 \\
    $s_{\text{init}}=0.005$    & 13.83                                   & 18.42                & 1.09                 \\
    $s_{\text{init}}=0.01$     & 14.21                                   & 19.05                & 1.09                 \\
    $s_{\text{init}}=0.05$     & 24.74                                   & 19.87                & 1.12                 \\
    $s_{\text{init}}=0.1$      & 25.23                                   & 28.32                & 1.15                 \\
    $s_{\text{init}}=0.2$      & 22.91                                   & 27.85                & 1.18                 \\
    \midrule
    \multicolumn{4}{l}{\underline{\textit{Scale constraints}}}                                                         \\
    Unconstrained              & 20.32                                   & 23.18                & 1.21                 \\
    Tight ($s \in [0.05,0.2]$) & 24.04                                   & 27.95                & 1.07                 \\
    Wide ($s \in [0.001,0.5]$) & 22.76                                   & 24.91                & 1.13                 \\
    \bottomrule
  \end{tabular}
\end{table}

The results indicate that \textcolor{stanfordred}{nGRF} is highly sensitive to Gaussian scaling parameters. For indoor environments, initial scaling values between 0.05 and 0.2 consistently yield good performance, while very small values perform poorly. The outdoor environment shows a preference for larger scaling values, with optimal performance observed when $s_{\text{init}}$ is between 0.1 and 0.2. This behavior is attributed to the nature of \textcolor{stanfordred}{nGRF}'s spatial weighting. Since the weight depends on the Gaussian's covariance, smaller Gaussians have a more localized influence. Achieving a good fit with small Gaussians would require a significantly larger number of them, which can lead to overfitting. Consequently, using larger Gaussians that cover more area and capture the overall field distribution is more effective.

Constraining scale parameters during training also improves performance, particularly when the constraints align with the optimal scale ranges for each environment. In these experiments, tight constraints centered around the optimal ranges ($s \in [0.05, 0.2]$) maintain high SNR while reducing variance. Unconstrained scales lead to performance degradation of approximately 4.9 dB indoors and 5.1 dB outdoors, as the Gaussians may converge to suboptimal scales.

\subsection{Positional encoding}

The frequency of the positional encoding affects the model's capacity to capture high-frequency variations in the electromagnetic field. Table~\ref{tab:pos_enc} shows the effect of using different numbers of frequency bands in the positional encoding on model performance.

\begin{table}[h]
  \centering
  \caption{\textit{Impact of positional encoding frequency bands.} Effect of the number of frequency bands on SNR and training time.}
  \label{tab:pos_enc}
  \small
  \setlength{\tabcolsep}{5pt}
  \renewcommand{\arraystretch}{1}
  \begin{tabular}{lcc}
    \toprule
    \textbf{Configuration} & \textbf{SNR (dB)}                       & \textbf{Train (min)} \\
    \midrule
    \cellcolor{highlightblue}\textbf{\textcolor{stanfordred}{nGRF} (baseline, $L=16$)}
                           & \cellcolor{highlightblue}\textbf{25.23}
                           & \cellcolor{highlightblue}\textbf{2.3}                          \\
    \midrule
    \multicolumn{3}{l}{\underline{\textit{Frequency bands}}}                                \\
    $L=4$                  & 21.07                                   & 2.0                  \\
    $L=8$                  & 23.83                                   & 2.1                  \\
    $L=10$                 & 23.95                                   & 2.1                  \\
    $L=12$                 & 24.05                                   & 2.2                  \\
    $L=16$                 & 25.23                                   & 2.3                  \\
    $L=32$                 & 23.35                                   & 2.4                  \\
    \bottomrule
  \end{tabular}
\end{table}

\begin{table}[t!]
  \centering
  \caption{\textit{Impact of network architecture on \textcolor{stanfordred}{nGRF} (conference).} Effect of depth, width, activations, and complex representation.}
  \label{tab:arch_ablation}
  \small
  \setlength{\tabcolsep}{2pt}
  \renewcommand{\arraystretch}{1}
  \begin{tabular}{lccc}
    \toprule
    \textbf{Configuration}         & \textbf{SNR (dB)}                       & \textbf{Params (M)} & \textbf{Train (min)} \\
    \midrule
    \multicolumn{4}{l}{\underline{\textit{\texttt{AttributeNetwork} depth}}}                                              \\
    2 layers (32-32)               & 21.45                                   & 0.8                 & 1.9                  \\
    3 layers (64-64-64)            & 24.18                                   & 1.2                 & 2.1                  \\
    \cellcolor{highlightblue}4 layers (128-128-128-128)
                                   & \cellcolor{highlightblue}\textbf{25.23}
                                   & \cellcolor{highlightblue}\textbf{1.8}
                                   & \cellcolor{highlightblue}\textbf{2.3}                                                \\
    5 layers (128$\times$5)        & 24.97                                   & 2.3                 & 2.8                  \\
    \midrule
    \multicolumn{4}{l}{\underline{\textit{\texttt{DecoderNetwork} width}}}                                                \\
    Hidden dim = 32                & 22.34                                   & 1.1                 & 2.2                  \\
    \cellcolor{highlightblue}Hidden dim = 64
                                   & \cellcolor{highlightblue}\textbf{25.23}
                                   & \cellcolor{highlightblue}\textbf{1.8}
                                   & \cellcolor{highlightblue}\textbf{2.3}                                                \\
    Hidden dim = 128               & 25.41                                   & 3.2                 & 2.7                  \\
    \midrule
    \multicolumn{4}{l}{\underline{\textit{Activation functions}}}                                                         \\
    ReLU (both)                    & 23.12                                   & 1.8                 & 2.3                  \\
    \cellcolor{highlightblue}Leaky ReLU (Attr.) + ReLU (Dec.)
                                   & \cellcolor{highlightblue}\textbf{25.23}
                                   & \cellcolor{highlightblue}\textbf{1.8}
                                   & \cellcolor{highlightblue}\textbf{2.3}                                                \\
    GELU (both)                    & 24.87                                   & 1.8                 & 2.4                  \\
    Swish (both)                   & 24.45                                   & 1.8                 & 2.4                  \\
    \midrule
    \multicolumn{4}{l}{\underline{\textit{Complex representation}}}                                                       \\
    \cellcolor{highlightblue}Separate Re/Im outputs
                                   & \cellcolor{highlightblue}\textbf{25.23}
                                   & \cellcolor{highlightblue}\textbf{1.8}
                                   & \cellcolor{highlightblue}\textbf{2.3}                                                \\
    Magnitude + phase              & 22.78                                   & 1.8                 & 2.4                  \\
    Polar coordinates $(r,\theta)$ & 21.94                                   & 1.8                 & 2.5                  \\
    \bottomrule
  \end{tabular}
\end{table}

The experiments show that \textcolor{stanfordred}{nGRF}'s performance is relatively insensitive to the specific number of positional encoding frequency bands, provided it exceeds a minimum threshold. When the number of frequency bands is very low ($L=4$), a performance decrease of 4.16 dB compared to the baseline is observed, indicating insufficient capacity to represent high-frequency spatial variations.

Beyond this threshold, however, performance remains remarkably stable. The SNR varies by at most 1 to 2 dB across all configurations from $L=8$ to $L=32$, with no clear monotonic improvement as the number of frequency bands increases. Performance peaks at $L=16$ and declines slightly with $L=32$, suggesting that excessive frequency bands may introduce unnecessary high-frequency components that lead to overfitting.

\subsection{Effect of network architecture}

Table~\ref{tab:arch_ablation} analyzes the impact of architectural choices in both the \texttt{AttributeNetwork} and \texttt{DecoderNetwork}. Increasing the depth of the \texttt{AttributeNetwork} improves performance up to four layers, achieving the best SNR of 25.23 dB, while deeper configurations offer no further gains and incur higher training cost, indicating diminishing returns. For the \texttt{DecoderNetwork}, a hidden dimension of 64 provides the most favorable accuracy-efficiency trade-off; increasing the width to 128 yields only marginal SNR improvement at the expense of a substantially larger parameter count and longer training time. Activation function ablations show that using Leaky ReLU in the \texttt{AttributeNetwork} combined with ReLU in the \texttt{DecoderNetwork} consistently outperforms symmetric choices such as ReLU, GELU, or Swish in both networks. Finally, representing the complex channel using separate real and imaginary outputs achieves the highest SNR, outperforming magnitude-phase and polar representations, which suffer from optimization instability due to phase discontinuities.

\subsection{Measurement density and noise}

This subsection analyzes how various dataset characteristics affect \textcolor{stanfordred}{nGRF}'s performance, focusing on measurement density and robustness to noise. Table~\ref{tab:dataset} summarizes these findings.

\begin{table}[t]
  \centering
  \caption{\textit{Impact of dataset characteristics.} Effect of measurement density and noise on SNR and training time.}
  \label{tab:dataset}
  \small
  \setlength{\tabcolsep}{6pt}
  \renewcommand{\arraystretch}{1}
  \begin{tabular}{lcc}
    \toprule
    \textbf{Configuration}      & \textbf{SNR (dB)}                       & \textbf{Train (min)} \\
    \midrule
    \cellcolor{highlightblue}\textbf{\textcolor{stanfordred}{nGRF} (baseline)}
                                & \cellcolor{highlightblue}\textbf{25.23}
                                & \cellcolor{highlightblue}\textbf{2.3}                          \\
    \midrule
    \multicolumn{3}{l}{\underline{\textit{Measurement density (samples/ft$^3$)}}}                \\
    0.005                       & 19.47                                   & 2.0                  \\
    0.01                        & 24.56                                   & 2.2                  \\
    0.02                        & 25.01                                   & 2.4                  \\
    0.05                        & 24.51                                   & 2.7                  \\
    \midrule
    \multicolumn{3}{l}{\underline{\textit{Noise level}}}                                         \\
    No noise ($\sigma=0$)       & 21.34                                   & 2.2                  \\
    Low noise ($\sigma=0.001$)  & 24.85                                   & 2.2                  \\
    Baseline ($\sigma=0.00387$) & 25.23                                   & 2.3                  \\
    Med.\ noise ($\sigma=0.02$) & 23.67                                   & 2.4                  \\
    High noise ($\sigma=0.1$)   & 18.42                                   & 2.8                  \\
    \bottomrule
  \end{tabular}
\end{table}

With just 0.01 samples/ft$^3$, \textcolor{stanfordred}{nGRF} achieves an SNR of 24.56 dB, only 0.67 dB below the baseline performance. Doubling the density to 0.02 samples/ft$^3$ yields only a minor improvement, while further increases result in a slight, negligible degradation in SNR. This result suggests that the structured Gaussian representation effectively interpolates between sparse measurements, allowing it to achieve high fidelity with 18$\times$ fewer measurements than comparable neural field approaches.

Consistent with prior work in neural fields, a certain level of noise during training is found to be beneficial. Training without noise ($\sigma=0$) results in an SNR 3.89 dB below the baseline. A small amount of positional noise ($\sigma=0.00387$ in the baseline) acts as a regularizer, preventing overfitting and encouraging the model to learn smoother, more generalizable field representations.

Excessive noise, however, is problematic. Medium noise levels ($\sigma=0.02$) reduce performance by 1.56 dB, while high noise ($\sigma=0.1$) causes a significant degradation of 6.81 dB. This highlights the existence of an optimal noise level and underscores the model's sensitivity to this hyperparameter, which is a notable limitation of \textcolor{stanfordred}{nGRF}.




\end{document}